\documentclass[english,aps,prd,notitlepage,10pt,superscriptaddress,preprintnumbers]{revtex4-2}
\usepackage[T1]{fontenc}
\usepackage[latin9]{inputenc}
\usepackage{amsmath}
\usepackage{graphicx}

\makeatletter
\def\Mpl{M_{\rm P}}

\makeatother

\usepackage{babel}
\begin{document}
\preprint{YITP-20-145, IPMU20-0116}

\title{Weakening gravity for dark matter in a type-II minimally modified gravity}
\author{Antonio De Felice}
\affiliation{Center for Gravitational Physics, Yukawa Institute for Theoretical
Physics, Kyoto University, 606-8502, Kyoto, Japan}
\email{antonio.defelice@yukawa.kyoto-u.ac.jp}

\author{Shinji Mukohyama}
\affiliation{Center for Gravitational Physics, Yukawa Institute for Theoretical
Physics, Kyoto University, 606-8502, Kyoto, Japan}
\affiliation{Kavli Institute for the Physics and Mathematics of the Universe (WPI),
The University of Tokyo, Kashiwa, Chiba 277-8583, Japan}
\email{shinji.mukohyama@yukawa.kyoto-u.ac.jp}

\date{\today}
\begin{abstract}
We propose a new cosmological framework in which the strength of the gravitational force acted on dark matter at late time can be weaker than that on the standard matter fields without introducing extra gravitational degrees of freedom. The framework integrates dark matter into a type-II minimally modified gravity that was recently proposed as a dark energy mimicker. The idea that makes such a framework possible consists of coupling a dark matter Lagrangian and a cosmological constant to the metric in a canonically transformed frame of general relativity (GR). On imposing a gauge fixing constraint, which explicitly breaks the temporal diffeomorphism invariance, we keep the number of gravitational degrees of freedom to be two, as in GR. We then make the inverse canonical transformation to bring the theory back to the original frame, where one can add the standard matter fields. This framework contains two free functions of time which specify the generating functional of the above mentioned canonical transformation and which are then used in order to realize desired time evolutions of both the Hubble expansion rate $H(z)$ and the effective gravitational constant for dark matter $G_{\rm eff}(z)$. The aim of this paper is therefore to provide a new framework to address the two puzzles present in today's cosmology, i.e.\ the $H_0$ tension and the $S_8$ tension, simultaneously. When the dark matter is cold in this framework, we dub the corresponding cosmological model the V Canonical Cold Dark Matter (VCCDM), as the cosmological constant $\Lambda$ in the standard $\Lambda$CDM is replaced by a function $V(\phi)$ of an auxiliary field $\phi$ and the CDM is minimally coupled to the metric in a canonically transformed frame. 
\end{abstract}
\maketitle

\section{Introduction}

In the last few years, there has been increasing evidence that cosmology might be revealing some unexpected new piece of the puzzle representing our knowledge of our universe. In particular there has been growing evidence that some tension is present among several different experimental results, in particular for the value of today's Hubble parameter $H_0$ (\cite{Aghanim:2018eyx,Riess:2019cxk,Bernal:2016gxb,Riess:2020sih,Wong:2019kwg,Reid:2008nm,Freedman:2019jwv}) and for the apparently lower growth rate seen in the data compared to the value predicted by $\Lambda$CDM \cite{Asgari:2020wuj,Said:2020epb,Boruah:2019icj}. As a consequence something might be wrong either with the data (but this assumption, if not supported by evidence, tends to lead to the impression that any physical attempt to make a theory for our universe would not be testable), or with the theory we have assumed, so far, to be the one able to describe everything which happens at the classical level. In order to check whether the data have some problems (mostly, yet unknown systematics), other and new experiments have been prepared to follow up previous experiments.  Following this path the Planck data have been recently confirmed by ATC \cite{Aiola:2020azj}, and local measurements of the Hubble parameter \cite{Riess:2019cxk}, $H_{0}$, have also been confirmed by independent measurements (e.g.\ \cite{Wong:2019kwg}). On top of this now very well known discrepancy between $\Lambda$CDM predictions and observational data, a probably less dramatic tension is related with the structure-formation. In this case data seem to prefer an effective gravitational constant which looks weaker than the one predicted in the context of the $\Lambda$CDM model in general relativity (GR) (see e.g.\ \cite{Asgari:2020wuj}).

Are all these observations telling us that we are simply not aware of all the systematics in our experiments, or rather that the $\Lambda$CDM model needs to be replaced with some other new theory of gravity (including possibly unknown dark energy or/and dark matter components)? Given the embarrassing status of the tension between theory and experiments, it seems that one needs to give an answer (whatever it is) to this question, which, on the other hand, cannot be dodged for much longer.

We thus adopt the assumption that the results obtained by observational data are trustable and, as a direct consequence, propose to change gravity and its interaction with dark matter in a minimal fashion. The aim is to address all the previously introduced tensions in the data with one single model. The idea to be explored in this paper follows after modified gravity theories introduced recently in the literature and called type-II minimally modified gravity (MMG), i.e.\ metric theories of gravity without an Einstein frame and endowed with only two gravitational degrees of freedom, e.g.\ \cite{DeFelice:2015hla,DeFelice:2015moy,DeFelice:2016ufg,DeFelice:2020eju,Bolis:2018vzs,DeFelice:2018vza,DeFelice:2020eju,DeFelice:2020cpt,DeFelice:2020onz,Aoki:2020lig,Aoki:2020iwm,Aoki:2020ila,Yao:2020tur}. In particular, the model introduced in \cite{DeFelice:2020eju,DeFelice:2020cpt,DeFelice:2020onz}, dubbed as VCDM, contains a time-dependent dark energy component, which due to the breaking of four-dimensional diffeomorphism at large scales is able to reproduce a given form of the Hubble expansion rate $H(z)$ without introducing any new physical degrees of freedom or instabilities (at the fully nonlinear level).

In this paper we propose a new cosmological framework to integrate the dark matter sector into the VCDM class of type-II MMG, in which a canonical transformation is characterized by two free functions $f_0(\phi)$ and $f_1(\phi)$ of an auxiliary field $\phi$. While in the original VCDM only one particular combination of the two free functions denoted as $V(\phi)$ enters the action and the equations of motion, in the new framework both of the two free functions are relevant. In the original VCDM a canonical transformation of GR to a new frame is generated by a generating functional characterized by $f_0(\phi)$ and $f_1(\phi)$ but only a cosmological constant $\bar{\Lambda}$ is introduced in the canonically transformed frame. As a result, only the combination $V(\phi)=\bar{\Lambda}f_1^{-1/2}f_0^{-3/2}$ affects the phenomenology of the theory. On the other hand, in the new framework studied in the present paper, not only the cosmological constant but also a dark matter Lagrangian is introduced in the canonically transformed frame of GR. As a result, we are now able to make use of both of these two free functions to affect two desired physical quantities. In particular, on focusing on the above mentioned $H_0$ and $S_8$ tensions, it is clear that we can use these two functions to reconstruct a given and desired dynamics of both $H(z)$ and $G_{\rm eff}(z)/G_N$. This framework thus opens up a new possibility to address the tensions between experimental data and theory with a minimal cost, in the sense that in addition to the cosmological constant and the dark matter (here dubbed as canonical cold dark matter (CCDM)) introduced in the canonically transformed frame, we do not add any new physical degrees of freedom nor new interactions to the theory. This fact makes it much easier to avoid ghost and gradient instabilities compared to, say, Horndeski, beyond Horndeski or DHOST models \cite{Horndeski:1974wa,Gleyzes:2014dya,Langlois:2017mxy}. In this sense, this approach is also similar to other recently proposed minimal modifications of gravity \cite{Lin:2017oow,Aoki:2018zcv,Mukohyama:2019unx,Aoki:2020oqc,Aoki:2018brq} (see also \cite{Afshordi:2006ad,Iyonaga:2018vnu,Feng:2019dwu,Gao:2019twq}). 

The rest of the present paper is organized as follows. Sec.~\ref{sec:model} provides a description of the model and the extension to a general framework in which the strength of the gravitational force acted on dark matter at late time can be weaker than that on the standard matter fields without introducing extra gravitational degrees of freedom. We first construct the Hamiltonian of the model by extending the construction of the VCDM class of type-II MMG and confirm that there is no extra physical degrees of freedom in the system. By a Legendre transformation we then obtain the Lagrangian of the model. We also provide a more general framework that can accommodate arbitrary model of dark matter with a covariant action. In Sec.~\ref{sec:reconstruction} we study the dynamics of a flat Friedmann-Lema\^{i}tre-Robertson-Walker (FLRW) background and develop a method to reconstruct the two free functions of time from cosmological observables. We then study perturbation equations of motion in Sec.~\ref{sec:perturbation}. In particular, restricting our consideration to a pressureless fluid in the standard matter sector and another pressureless fluid in the dark sector, we compute the effective gravitational constants in the model as a $2\times 2$ matrix. Among the four components of the matrix, those two that represent the strengths of gravity acted on the standard matter are constant and agree with the Newton's constant. On the other hand, the remaining two that represent the strengths of gravity acted on dark matter (and that happen to agree with each other) are time-dependent and can be smaller than the Newton's constant at late time. Sec.~\ref{sec:concl} is then devoted to a summary of this paper.

\section{Model description}
\label{sec:model}

In the context of MMG, in a recent paper \cite{DeFelice:2020eju}, it was introduced a new model which has the property of giving an effective time-dependent cosmological constant without introducing any new degree of freedom with respect to GR. This last fact is of importance and makes it possible for this model to recover GR at solar system scales without implementing an additional mechanism. In other words, not having extra scalar degrees of freedom, we do not need a Chameleon or Vainshtein mechanism in order to suppress any fifth force coming from extra degrees of freedom, as it usually happens instead in scalar-tensor theories. 

Here we want to further extend the theory mentioned above as to integrate the dark matter sector into the formalism so that the strength of gravitational force acted on dark matter is controllable without changing that on the standard matter fields. The motivation for such a new framework consists mainly in the possibility for the dark matter sector to feel a weaker gravity compared to the standard matter fields.

\subsection{Hamiltonian}

We start introducing the model by its Hamiltonian. The initial steps for defining the model coincide with the same steps of the MMG theory \cite{DeFelice:2020eju}. We start from the GR Hamiltonian
\begin{equation}
H_{{\rm GR}}=\int d^{3}x\,[N\mathcal{H}_{0}(\gamma,\pi)+N^{i}\mathcal{H}_{i}(\gamma,\pi)+\lambda\pi_{N}+\lambda^{i}\pi_{i}]\,,
\end{equation}
where 
\begin{eqnarray*}
\mathcal{H}_{0} & = & \frac{2}{\Mpl^{2}\sqrt{\gamma}}\left(\gamma_{ik}\gamma_{jl}-\frac{1}{2}\,\gamma_{ij}\gamma_{kl}\right)\pi^{ij}\pi^{kl}-\frac{\Mpl^{2}\sqrt{\gamma}}{2}\,R(\gamma)\,,\\
\mathcal{H}_{i} & = & -2\sqrt{\gamma}\gamma_{ij}D_{k}\!\left(\frac{\pi^{jk}}{\sqrt{\gamma}}\right).
\end{eqnarray*}
We then perform a canonical transformation via the following generating functional, where a tilde indicates new variables. 
\begin{equation}
F(\tilde{N},\tilde{N}^{i},\tilde{\gamma}_{ij},\pi_{N},\pi_{i},\pi^{ij}) 
 = -\int d^{3}x\,[\Mpl^{2}\sqrt{\tilde{\gamma}}f(\phi,\psi)+\tilde{N}^{i}\,\pi_{i}]\,,
\end{equation}
where $\phi$ and $\psi$ represent two 3D scalar combinations:
\begin{equation}
\phi=\frac{1}{\Mpl^{2}\sqrt{\tilde{\gamma}}}\,\pi^{ij}\,\tilde{\gamma}_{ij}\,,\qquad\psi=\frac{1}{\Mpl^{2}\sqrt{\tilde{\gamma}}}\,\pi_{N}\,\tilde{N}\,.
\end{equation}
The above introduced generation functional defines the following transformation equations:
\begin{eqnarray}
\tilde{\pi}_{\tilde{N}} & = & -\frac{\delta F}{\delta\tilde{N}}=\Mpl^{2}\sqrt{\tilde{\gamma}}f_{1}\,\frac{1}{\Mpl^{2}\sqrt{\tilde{\gamma}}}\,\pi_{N}=f_{1}\,\pi_{N}\,,\\
\tilde{\pi}_{i} & = & -\frac{\delta F}{\delta\tilde{N}^{i}}=\pi_{i}\,,\\
\tilde{\pi}^{ij} & = & -\frac{\delta F}{\delta\tilde{\gamma}_{ij}}=f_{0}\,\pi^{ij}+\frac{\Mpl^{2}}{2}\,\sqrt{\tilde{\gamma}}\,\tilde{\gamma}^{ij}\,(f-f_{0}\phi-f_{1}\psi)\,,\\
\gamma_{ij} & = & -\frac{\delta F}{\delta\pi^{ij}}=\Mpl^{2}\sqrt{\tilde{\gamma}}f_{0}\,\frac{1}{\Mpl^{2}\sqrt{\tilde{\gamma}}}\,\tilde{\gamma}_{ij}=f_{0}\,\tilde{\gamma}_{ij}\,,\\
N & = & -\frac{\delta F}{\delta\pi_{N}}=f_{1}\,\tilde{N}\,,\\
N^{i} & = & -\frac{\delta F}{\delta\pi_{i}}=\tilde{N}^{i}\,,
\end{eqnarray}
where we have defined $f_{0}\equiv \partial f/\partial\phi$ and $f_{1}\equiv \partial f/\partial\psi$. Therefore we can also write
\begin{eqnarray}
\psi & = & \frac{1}{\Mpl^{2}\sqrt{\tilde{\gamma}}}\,\pi_{N}\,\tilde{N}=\frac{f_{0}^{3/2}}{\Mpl^{2}\sqrt{\gamma}}\,\frac{\pi_{N}}{f_{1}}\,N\,,\label{eq:constr_psi}\\
\phi & = & \frac{1}{\Mpl^{2}\sqrt{\tilde{\gamma}}}\,\pi^{ij}\,\tilde{\gamma}_{ij}=\frac{f_{0}^{3/2}}{\Mpl^{2}f_{0}\sqrt{\gamma}}\,\pi^{ij}\,\gamma_{ij}=\frac{f_{0}^{1/2}}{\Mpl^{2}\sqrt{\gamma}}\,\pi^{ij}\,\gamma_{ij}\,.\label{eq:constr_phi}
\end{eqnarray}

After performing the canonical transformation, we promote $\phi$ and $\psi$ to be two independent 3D scalars by imposing (\ref{eq:constr_psi}) and (\ref{eq:constr_phi}) as constraints. We then introduce a cosmological constant $\bar{\Lambda}$ (in the new frame), a gauge fixing term (to keep the degrees of freedom to be only two), and two Lagrange multipliers which are meant to set the constraints (\ref{eq:constr_psi}) and (\ref{eq:constr_phi}). We then have the Hamiltonian of the MMG in the canonically transformed frame as 
\begin{eqnarray}
 H_{{\rm MMG}} & = & \int d^{3}x[\tilde{N}f_{1}\mathcal{H}_{0}(\tilde{\gamma},\tilde{\pi},\phi,\psi)+\tilde{N}^{i}\mathcal{H}_{i}(\tilde{\gamma},\tilde{\pi},\phi,\psi)+\tilde{\lambda}\tilde{\pi}_{\tilde{N}}+\lambda^{i}\tilde{\pi}_{i}+\sqrt{\gamma}\lambda_{C}\mathcal{C}(\tilde{\gamma},\tilde{\pi},\phi,\psi)\nonumber \\
 &  & {}+\sqrt{\gamma}\lambda_{D}\mathcal{D}(\tilde{\gamma},\tilde{\pi},\phi,\psi)+\lambda_{\phi}\pi_{\phi}+\lambda_{\psi}\pi_{\psi}+\sqrt{\gamma}\lambda_{{\rm gf}}^{i}\partial_{i}\phi+\Mpl^{2}\tilde{N}\sqrt{\tilde{\gamma}}\tilde{\Lambda}]\,.
\end{eqnarray}

At this level we are ready to introduce the action of dark matter. A general framework will be given in subsection~\ref{subsec:generalframework}. In the rest of the present paper including this subsection, for simplicity we adopt the low energy effective description of cold dark matter (CDM) as a pressureless fluid. (See Appendix for a brief review of the description in GR.) Then we add the CDM Hamiltonian written in the tilde-metric variables, as follows
\begin{eqnarray}
H_{{\rm tot}} & = & \int d^{3}x[\tilde{N}f_{1}\mathcal{H}_{0}(\tilde{\gamma},\tilde{\pi},\phi,\psi)+\tilde{N}^{i}\mathcal{H}_{i}(\tilde{\gamma},\tilde{\pi},\phi,\psi)+\tilde{\lambda}\tilde{\pi}_{\tilde{N}}+\lambda^{i}\tilde{\pi}_{i}+\sqrt{\gamma}\lambda_{C}\mathcal{C}(\tilde{\gamma},\tilde{\pi},\phi,\psi)\nonumber \\
 &  & {}+\sqrt{\gamma}\lambda_{D}\mathcal{D}(\tilde{\gamma},\tilde{\pi},\phi,\psi)+\lambda_{\phi}\pi_{\phi}+\lambda_{\psi}\pi_{\psi}+\sqrt{\gamma}\lambda_{{\rm gf}}^{i}\partial_{i}\phi+\Mpl^{2}\tilde{N}\sqrt{\tilde{\gamma}}\tilde{\Lambda}\nonumber \\
 &  & {}+\frac{\tilde{N}\sqrt{\tilde{\gamma}}}{2\rho}\left[\frac{\pi_{\sigma}^{2}}{\tilde{\gamma}}+\rho^{2}\left(1+\partial_{i}\sigma\partial_{j}\sigma\,\tilde{\gamma}^{ij}\right)\right]+\lambda_{\rho}\pi_{\rho}+\pi_{\sigma}\tilde{N}^{i}\partial_{i}\sigma+\lambda_{2}\,\sqrt{\tilde{\gamma}}\left(1-\frac{\pi_{\sigma}^{2}}{\rho^{2}\tilde{\gamma}}+\partial_{i}\sigma\partial_{j}\sigma\,\tilde{\gamma}^{ij}\right),
\end{eqnarray}
where the last line corresponds to the action of a dust fluid written in the tilded-new-frame.

On using the inverse canonical transformation, we find the Hamiltonian in terms of the original frame as 
\begin{eqnarray}
H_{{\rm tot}} & = & \int d^{3}x\left\{N\mathcal{H}_{0}(\gamma,\pi)+N^{i}\mathcal{H}_{i}(\gamma,\pi)+\lambda\,\pi_{N}+\lambda^{i}\pi_{i}+\sqrt{\gamma}\lambda_{C}\left(\phi-\frac{f_{0}^{1/2}}{\Mpl^{2}}\,\frac{\pi^{ij}}{\sqrt{\gamma}}\,\gamma_{ij}\right)\right.\nonumber \\
 &  & {}+\sqrt{\gamma}\lambda_{D}\left(\psi-\frac{f_{0}^{3/2}}{\Mpl^{2}f_{1}}\,\frac{\pi_{N}}{\sqrt{\gamma}}\,N\right)+\lambda_{\phi}\pi_{\phi}+\lambda_{\psi}\pi_{\psi}+\sqrt{\gamma}\,\lambda_{{\rm gf}}^{i}\,\partial_{i}\phi+\frac{\Mpl^{2}}{f_{1}f_{0}^{3/2}}\,N\sqrt{\gamma}\tilde{\Lambda}\nonumber \\
 &  & {}+\lambda_{\rho}\pi_{\rho}+\pi_{\sigma}N^{i}\partial_{i}\sigma+\frac{N\sqrt{\gamma}}{2\rho f_{0}^{3/2}f_{1}}\left[\rho^{2}\left(1+f_{0}\partial_{i}\sigma\partial_{j}\sigma\,\gamma^{ij}\right)+f_{0}^{3}\,\frac{\pi_{\sigma}^{2}}{\gamma}\right]\nonumber\\
& & {}+\left.\frac{\lambda_{2}}{f_{0}^{3/2}}\,\sqrt{\gamma}\left(f_{0}\partial_{i}\sigma\partial_{j}\sigma\,\gamma^{ij}+1-\frac{f_{0}^{3}\pi_{\sigma}^{2}}{\rho^{2}\gamma}\right)\right\}.
\end{eqnarray}
The following two primary constraints 
\begin{eqnarray}
\pi_{N} & \approx & 0\,,\\
\psi & \approx & \frac{f_{0}^{3/2}}{\Mpl^{2}f_{1}}\,\frac{\pi_{N}}{\sqrt{\gamma}}\,N\,,
\end{eqnarray}
automatically lead to the extra constraint
\begin{equation}
\psi\approx0\,.
\end{equation}
We also have another primary constraint
\begin{equation}
\pi_{\psi}\approx0\,,
\end{equation}
so that we can eliminate the pair $(\psi,\pi_{\psi})$ from the dynamical variables. In this case we can expand $f$ as a function of $\phi$ and $\psi$, as in 
\begin{equation}
f(\phi,\psi)=F(\phi)+f_{1}(\phi)\,\psi+\mathcal{O}(\psi^{2})\,,
\end{equation}
and stop at the linear order, as any higher order terms would disappear from the Hamiltonian and the constraint algebra, because of the constraint $\psi\approx0$. 

Furthermore, in the following we consider $N$ and $N^{i}$ as Lagrange multipliers. Finally, on making a field redefinition for the fields
$\phi$, $\lambda_{C}$, and $\lambda_{2}$, we find
\begin{eqnarray}
H_{{\rm tot}} & = & \int d^{3}x\left[N\mathcal{H}_{0}(\gamma,\pi)+N^{i}\mathcal{H}_{i}(\gamma,\pi)+\sqrt{\gamma}\lambda_{C}\left(\Mpl^{2}\phi-\frac{\pi^{ij}}{\sqrt{\gamma}}\,\gamma_{ij}\right)\right.\nonumber \\
 &  & {}+\lambda_{\phi}\pi_{\phi}+\sqrt{\gamma}\,\lambda_{{\rm gf}}^{i}\,\partial_{i}\phi+\Mpl^{2}\,N\sqrt{\gamma}\,V(\phi)+\lambda_{\rho}\pi_{\rho}+\pi_{\sigma}N^{i}\partial_{i}\sigma\nonumber \\
 &  & {}+\left.\frac{N\sqrt{\gamma}\,\rho}{2f_{1}f_{0}^{3/2}}\left(1+f_{0}\partial_{i}\sigma\partial_{j}\sigma\,\gamma^{ij}+\frac{f_{0}^{3}\,\pi_{\sigma}^{2}}{\gamma\rho^{2}}\right)+\lambda_{2}\,\sqrt{\gamma}\left(1+f_{0}\partial_{i}\sigma\partial_{j}\sigma\,\gamma^{ij}-\frac{f_{0}^{3}\pi_{\sigma}^{2}}{\gamma\rho^{2}}\right)\right],\label{eq:CCDM_ham}
\end{eqnarray}
where we have redefined $\phi$, $\lambda_{C}$ and $\lambda_2$, and have defined $V(\phi)\equiv\bar{\Lambda}/(f_1f_0^{3/2})$.

\subsection{Number of degrees of freedom}

We can see that the Hamiltonian of Eq.\ (\ref{eq:CCDM_ham}) implements the following constraints
\begin{eqnarray}
C_{1} & \equiv & \mathcal{H}_{0}+\Mpl^{2}\,\sqrt{\gamma}\,\Lambda(\phi)+\frac{\sqrt{\gamma}\,\rho}{2f_{1}f_{0}^{3/2}}\left(1+\beta(\phi)\partial_{i}\sigma\partial_{j}\sigma\,\gamma^{ij}+\frac{\beta^{3}\,\pi_{\sigma}^{2}}{\gamma\rho^{2}}\right),\\
C_{2} & \equiv & \sqrt{\gamma}\bar{\pi}_{\phi}\,,\\
C_{3} & \equiv & \sqrt{\gamma}\left(\Mpl^{2}\phi-\frac{\pi^{ij}}{\sqrt{\gamma}}\,\gamma_{ij}\right),\\
C_{4,i}^{{\rm ext}} & \equiv & \sqrt{\gamma}\left(\bar{\pi}_{\sigma}\partial_{i}\sigma+\bar{\pi}_{\phi}\partial_{i}\phi+\bar{\pi}_{\rho}\partial_{i}\rho-2\gamma_{ik}\nabla_{j}\bar{\pi}^{kj}\right),\\
C_{5,i} & \equiv & \sqrt{\gamma}\,\partial_{i}\phi\,,\qquad\textrm{or equivalently}\qquad C_{5}=\sqrt{\gamma}\nabla^{2}\phi\,,\\
C_{6} & \equiv & \sqrt{\gamma}\bar{\pi}_{\rho}\,,\\
C_{7} & \equiv & 1+\beta\partial_{i}\sigma\partial_{j}\sigma\,\gamma^{ij}-\frac{\beta^{3}\pi_{\sigma}^{2}}{\gamma\rho^{2}}\,.
\end{eqnarray}
It can be proven that $\{\int v^{i}C_{4,i}^{{\rm ext}},H_{{\rm tot}}\}_{{\rm PB}}\approx0$ is automatically satisfied on the constraint surface, so it does not fix Lagrange multipliers or add any new constraint. The relation $\{\int v^{i}C_{5,i},H_{{\rm tot}}\}_{{\rm PB}}\approx0$ can be solved for $\lambda_{\phi}$, whereas $\{\int vC_{6},H_{{\rm tot}}\}_{{\rm PB}}\approx0$ gives $\lambda_{2}\approx0$. The constraint $\{\int vC_{2},H_{{\rm tot}}\}_{{\rm PB}}\approx0$ can be solved for $\nabla_{i}\lambda_{{\rm gf}}^{i}$. The constraint $\{\int vC_{7},H_{{\rm tot}}\}_{{\rm PB}}\approx0$ sets the Lagrange multiplier $\lambda_{\rho}$. The time derivative of the Hamiltonian constraint can be solved for $\lambda_{C}$, and, finally $\{\int vC_{3},H_{{\rm tot}}\}_{{\rm PB}}\approx0$ can be used to fix $N$.

In can also be proven that the vector density $C_{4,i}^{{\rm ext}}$ consists of 3 first-class constraints, as they have vanishing Poisson brackets with any other constraint. All other constraints are of second-class. Therefore we can now deduce the number of degrees of freedom. In fact, on the gravity side, on considering $N$ and $N^{i}$ as Lagrange multipliers, we get $\gamma_{ij}$ so $6\times2$ variables and $3\times2$ variables coming from the 3 scalar fields, $\phi$, $\rho$ and $\sigma$. So in total we have 18 variables in the phase space that we started with, $3$ first-class constraints and $6$ second-class constraints (where we have counted $C_{5,i}$ as one, as it can be rewritten as $\lambda_{{\rm gf}}\nabla^{2}\phi$). Therefore we are left with $\frac{1}{2}(18-3\times 2 -6)=3$ degrees of freedom, as expected. Obviously, $2$ among $3$ represent tensorial gravitational waves, while the remaining $1$ represents the dark matter fluid. 

\subsection{Lagrangian}

The Lagrangian for this theory can be found via a Legendre transformation as follows. First of all
\begin{eqnarray}
\frac{\pi^{ij}}{\sqrt{\gamma}} & = & \frac{\Mpl^{2}}{2}\,(K^{ij}-\gamma^{ij}K)-\frac{\Mpl^{2}}{2}\,\frac{\lambda}{N}\,\gamma^{ij}\,,\\
\frac{\pi_{\sigma}}{\sqrt{\gamma}} & = & \frac{\rho^{2}\,(\dot{\sigma}-N^{i}\partial_{i}\sigma)}{\beta^{3}\,[N\,\rho/(f_{1}f_{0}^{3/2})-2\lambda_{2}]}\,,\\
\dot{\gamma}_{ij} & = & 2NK_{ij}+\nabla_{i}N_{j}+\nabla_{j}N_{i}\,,
\end{eqnarray}
and then we find the Lagrangian by direct computations. After we have the Lagrangian, we make a field redefinition as
\begin{equation}
\frac{N}{f_{1}f_{0}^{3/2}}\rho-2\lambda_{2}=\frac{\rho^{2}}{\lambda_{3}}\,,
\end{equation}
and then we find that the equation of motion for $\rho$ can be solved
for $\lambda_{3}$, giving
\begin{equation}
\lambda_{3}=\frac{\rho f_{1}f_{0}^{3/2}}{N\,}\,,
\end{equation}
or, equivalently $\lambda_{2}=0$. On plugging it into the Lagrangian we find
\begin{eqnarray}
\mathcal{L} & = & \frac{\Mpl^{2}}{2}\,N\sqrt{\gamma}\,[K_{ij}K^{ij}-K^{2}+R-2V(\phi)]-\frac{3\Mpl^{2}}{4}\,N\sqrt{\gamma}\,\lambda^{2}-\Mpl^{2}N\sqrt{\gamma}\,\lambda\,(K+\phi)-\sqrt{\gamma}\lambda_{{\rm gf}}^{i}\partial_{i}\phi\nonumber \\
 &  & {}+\frac{N\sqrt{\gamma}\rho}{2f_{1}f_{0}^{3/2}}\left[\frac{f_{1}^{2}\dot{\sigma}^{2}}{N^{2}}-\frac{2f_{1}^{2}\dot{\sigma}\,N^{i}\partial_{i}\sigma}{N^{2}}-1+\left(\frac{f_{1}^{2}N^{i}N^{j}}{N^{2}}-f_{0}\gamma^{ij}\right)\partial_{i}\sigma\partial_{j}\sigma\right]\,, \label{eqn:VCCDM}
\end{eqnarray}
where the second line corresponds to the CCDM Lagrangian. In the limit $f_{0}\to1$ and $f_{0}\to1$, we get back the standard action of CDM (see Appendix). On considering $N>0$, and $\tilde{N}>0$, we need to impose $f_{1}>0$, together also with $f_{0}>0$ as to keep $\tilde{\gamma}_{ij}$ positive definite. 

Actually, we can immediately see that the CCDM Lagrangian can be written in the form of the standard Lagrangian of a pressureless fluid, provided that we introduce effective lapse and spatial metric variables as follows
\begin{eqnarray}
N_{{\rm eff}}^{2} & = & \frac{N^{2}}{f_{1}^{2}}\,,\\
\gamma^{{\rm eff}}_{ij} & = & \frac{\gamma_{ij}}{f_{0}}\,,
\end{eqnarray}
which lead to
\begin{eqnarray}
\mathcal{L} & = & \frac{\Mpl^{2}}{2}\,N\sqrt{\gamma}\,[K_{ij}K^{ij}-K^{2}+R-2V(\phi)]-\frac{3\Mpl^{2}}{4}\,N\sqrt{\gamma}\,\lambda^{2}-\Mpl^{2}N\sqrt{\gamma}\,\lambda\,(K+\phi)-\sqrt{\gamma}\lambda_{{\rm gf}}^{i}\nabla_{i}\phi+\mathcal{L}_{{\rm m}}\,,\\
\mathcal{L}_{{\rm m}} & = & \frac{N_{{\rm eff}}\sqrt{\gamma_{{\rm eff}}}\rho_{{\rm cdm}}}{2}\left[\frac{\dot{\sigma}^{2}}{N_{{\rm eff}}^{2}}-\frac{2\dot{\sigma}N^{i}\nabla_{i}\sigma}{N_{{\rm eff}}^{2}}-1+\left(\frac{N^{i}N^{j}}{N_{{\rm eff}}^{2}}-\gamma_{{\rm eff}}^{ij}\right)\nabla_{i}\sigma\nabla_{j}\sigma\right]\nonumber\\
 & = &-\frac{1}{2}\sqrt{-g^{{\rm eff}}}\,\rho_{{\rm cdm}}(\partial_{\mu}\sigma\,\partial_{\nu}\sigma\,g_{{\rm eff}}^{\mu\nu}+1)\,,
\end{eqnarray}
as expected by the canonical transformation relations, where
\begin{equation}
g^{\rm eff}_{\mu\nu}dx^{\mu}dx^{\nu} = -N_{\rm eff}^2dt^2 + \gamma^{\rm eff}_{ij}(dx^i+N^idt)(dx^j+N^jdt)\,,
\end{equation}
and $g_{\rm eff}^{\mu\nu}$ is the inverse of $g^{\rm eff}_{\mu\nu}$. Here, the field $\rho$ has been renamed as $\rho_{{\rm cdm}}$. 

Finally, we introduce the standard matter fields as being coupled to the original metric variables ($N$, $N^i$, $\gamma_{ij}$).

\subsection{General framework}
\label{subsec:generalframework}

In the previous subsections we have treated dark matter as a pressureless fluid, which is simply a low energy effective description of a more fundamental action. In this subsection we extend the setup developed so far to accommodate an arbitrary model of dark matter with a covariant action. 

Starting with the ADM decomposition of a $4$-dimensional spacetime metric $g_{\mu\nu}$ as
\begin{equation}
 g_{\mu\nu}dx^{\mu}dx^{\nu} = - N^2dt^2 + \gamma_{ij}(dx^i+N^idt)(dx^j+N^jdt)\,,  \label{eqn:4dmetric-visible}
\end{equation}
we introduce an auxiliary scalar field $\phi$ and choose two arbitrary positive functions $f_0(\phi)$ and $f_1(\phi)$ of $\phi$. We then introduce an effective metric for dark matter as
\begin{equation}
g^{\rm eff}_{\mu\nu}dx^{\mu}dx^{\nu} = - \frac{N^2}{f_1^2}dt^2 + \frac{\gamma_{ij}}{f_0}(dx^i+N^idt)(dx^j+N^jdt)\,.  \label{eqn:4dmetric-dark}
\end{equation}
The total action is then written as
\begin{equation}
 I = I_{\rm g}[N, N^i, \gamma_{ij},\phi,\lambda, \lambda_{\rm gf}^i] + I_{\rm sm}[g_{\mu\nu},\chi_{\rm sm}] + I_{\rm dm}[g^{\rm eff}_{\mu\nu},\chi_{\rm dm}]\,, \label{eqn:totalaction-general}
\end{equation}
where 
\begin{equation}
  I_{\rm g}[N, N^i, \gamma_{ij},\phi,\lambda, \lambda_{\rm gf}^i]
   = \frac{\Mpl^2}{2}\int dt d^3x N\sqrt{\gamma}
   \left[ K^{ij}K_{ij}-K^2 + R - 2V(\phi) -2(K+\phi)\lambda - \frac{3}{2}\lambda^2
    -\frac{2}{N}\lambda_{\rm gf}^i\nabla_i\phi \right] \,,
\end{equation}
$I_{\rm sm}[g_{\mu\nu},\chi_{\rm sm}]$ is the action of the standard matter including the standard model of particle physics coupled to the metric $g_{\mu\nu}$, and $I_{\rm dm}[g^{\rm eff}_{\mu\nu},\chi_{\rm dm}]$ is the action of dark matter coupled to the effective metric $g^{\rm eff}_{\mu\nu}$. Here, 
\begin{equation}
 V(\phi) \equiv \frac{\bar{\Lambda}}{f_1f_0^{3/2}}\,,
\end{equation}
and $\bar{\Lambda}$ is a constant.

For the standard matter coupled to the metric $g_{\mu\nu}$, we define its stress-energy tensor as usual,
\begin{equation}
 T^{\mu\nu} = \frac{2}{\sqrt{-g}}\frac{\delta I_{\rm sm}}{\delta g_{\mu\nu}}\,, \quad
  T^{\mu}_{\  \nu} = T^{\mu\rho}g_{\rho\nu}\,, \quad
  T_{\mu\nu} = T^{\rho\sigma}g_{\rho\mu}g_{\sigma\nu}\,. \label{eqn:Tmunu-sm}
\end{equation}
For the dark matter coupled to the effective metric $g^{\rm eff}_{\mu\nu}$, on the other hand, we define its stress-energy tensor through the functional derivative with respect to $g^{\rm eff}_{\mu\nu}$ as
\begin{equation}
 T_{\rm dm}^{\mu\nu} = \frac{2}{\sqrt{-g^{\rm eff}}}\frac{\delta I_{\rm dm}}{\delta g^{\rm eff}_{\mu\nu}}\,, \quad
  T_{{\rm dm}\, \nu}^{\ \mu} = T_{\rm dm}^{\mu\rho}g^{\rm eff}_{\rho\nu}\,, \quad
  T^{\rm dm}_{\mu\nu} = T_{{\rm dm}}^{\rho\sigma}g^{\rm eff}_{\rho\mu}g^{\rm eff}_{\sigma\nu}\,.
\end{equation}
For latter convenience we also define
\begin{equation}
 \bar{T}^{\mu\nu} = T_{\rm dm}^{\mu\nu} - \Mpl^2\bar{\Lambda}g_{\rm eff}^{\mu\nu}\,, \quad
 \bar{T}^{\mu}_{\ \nu} = T_{{\rm dm}\, \nu}^{\ \mu} - \Mpl^2\bar{\Lambda}\delta^{\mu}_{\ \nu}\,, \quad
 \bar{T}_{\mu\nu} = T^{\rm dm}_{\mu\nu} - \Mpl^2\bar{\Lambda}g^{\rm eff}_{\mu\nu}\,, 
\end{equation}
which is the total stress-energy tensor of the dark matter and the cosmological constant introduced in the canonically transformed frame. 

\section{Reconstruction of two free functions}
\label{sec:reconstruction}

In this section we develop a method to reconstruct the two free functions $f_0(\phi)$ and $f_1(\phi)$ from cosmological observables. We consider a flat FLRW metric
\begin{equation}
 N = N(t)\,, \quad N^i = 0\,, \quad \gamma_{ij}=a^2(t)\delta_{ij}\,,
\end{equation}
with 
\begin{equation}
 \phi = \phi(t)\,, \quad \lambda = \lambda(t)\,, \quad \lambda^i_{\rm gf}=0\,.
\end{equation}
As for the stress-energy tensors, we have
\begin{equation}
 T^{0}_{\ 0} = -\rho(t)\,, \quad T^{0}_{\ i} = 0\,, \quad T^{i}_{\ j} = P(t)\delta^{i}_{\ j}
\end{equation}
for the standard matter sector, and 
\begin{equation}
 \bar{T}^{0}_{\ 0} = -\bar{\rho}(t)\,, \quad \bar{T}^{0}_{\ i} = 0\,, \quad \bar{T}^{i}_{\ j} = \bar{P}(t)\delta^{i}_{\ j}
\end{equation}
for the dark sector. 

The scale factor $a(t)$ and the lapse function $N(t)$ in the standard matter frame are related to those in the dark frame, $\bar{a}(t)$ and $\bar{N}(t)$, as 
\begin{equation}
a(t)=f_{0}^{1/2}(\phi)\bar{a}(t)\,,\quad N(t)=f_{1}(\phi)\bar{N}(t)\,.
\end{equation}
As already stated, we assume that $f_{0}>0$ and $f_{1}>0$. The number of e-foldings $\mathcal{N}$ in the standard matter frame and that in the dark frame, $\bar{\mathcal{N}}$, are defined as 
\begin{equation}
d\mathcal{N}=HNdt\,,\quad H=\frac{1}{N}\frac{\dot{a}}{a}\,,
\end{equation}
and 
\begin{equation}
d\bar{\mathcal{N}}=\bar{H}\bar{N}dt\,,\quad\bar{H}=\frac{1}{\bar{N}}\frac{\dot{\bar{a}}}{\bar{a}}\,,
\end{equation}
and thus related to each other as 
\begin{equation}
\frac{d\bar{\mathcal{N}}}{d\mathcal{N}}=1-\frac{1}{2}\frac{d\ln f_{0}}{d\phi}\frac{d\phi}{d\mathcal{N}}\,.
\end{equation}
It is convenient to define the following combination 
\begin{equation}
\Theta=\frac{f_{0}^{\Gamma_{0}}}{f_{1}^{2\Gamma_{1}}}\,,
\end{equation}
where $\Gamma_0$ and $\Gamma_1$ are constants to be determined so that $\Theta$ corresponds to a cosmological observable by which we aim to set constraints on the theory. In the following we will choose the constants $\Gamma_{0},\Gamma_{1}$ as to make $\Theta$ coincident with $G_{{\rm eff}}/G_{N}$, but at the moment we leave them as free parameters.

Assuming that $H\ne0$, the set of background equations of motion is 
\begin{align}
\phi^{2} & =\frac{3}{M_{{\rm Pl}}^{2}}\left[\rho+\bar{\rho}v\right]\,,\label{eqn:V}\\
\frac{d\phi}{d\mathcal{N}} & =\frac{3}{2M_{{\rm Pl}}^{2}H}\left[(\rho+P)+(\bar{\rho}+\bar{P})v\right]\,,\label{eqn:dphidN}\\
\frac{d\rho_{i}}{d\mathcal{N}} & =-3(\rho_{i}+P_{i})\,,\\
\frac{d\bar{\rho}_{n}}{d\bar{\mathcal{N}}} & =-3(\bar{\rho}_{n}+\bar{P}_{n})\,,
\end{align}
where $\rho=\sum_{I}\rho_{I}$, $P=\sum_{I}P_{I}$, $\bar{\rho}=\sum_{n}\bar{\rho}_{n}$, $\bar{P}=\sum_{n}\bar{P}_{n}$ and we have introduced 
\begin{equation}
v\equiv\frac{1}{f_{0}^{3/2}f_{1}}\,,
\end{equation}
so that 
\begin{equation}
f_{0}=\Theta^{1/[\Gamma_{0}+3\Gamma_{1}]}v^{-2\Gamma_{1}/[\Gamma_{0}+3\Gamma_{1}]}\,,\quad f_{1}=\Theta^{-3/[2\Gamma_{0}+6\Gamma_{1}]}v^{-\Gamma_{0}/[\Gamma_{0}+3\Gamma_{1}]}\,.\label{eqn:f0f1-geffv}
\end{equation}
Here, $I$ runs over different components in the standard matter sector and $n$ runs over those in the dark sector. Provided that $H\ne0$, $\bar{\rho}\ne0$ and $(\rho+P)+(\bar{\rho}+\bar{P})v\ne0$, the following equations follow from the above equations. 
\begin{align}
v & =\frac{M_{{\rm Pl}}^{2}\phi^{2}-3\rho}{3\bar{\rho}}\,,\label{eqn:v}\\
\bar{\rho}\frac{dv}{d\phi} & =\frac{2}{3}M_{{\rm Pl}}^{2}\phi+2M_{{\rm Pl}}^{2}H\frac{(\rho+P)+(\bar{\rho}+\bar{P})v\frac{d\bar{\mathcal{N}}}{d\mathcal{N}}}{(\rho+P)+(\bar{\rho}+\bar{P})v}\,,\label{eqn:dv/dphi}\\
\frac{d\bar{\mathcal{N}}}{d\mathcal{N}} & =1-\frac{1}{2}\frac{1}{\Gamma_{0}+3\Gamma_{1}}\,\frac{d\ln\Theta}{d\mathcal{N}}+\frac{3\Gamma_{1}}{\Gamma_{0}+3\Gamma_{1}}\,\frac{dv}{d\phi}\,\frac{(\rho+P)+(\bar{\rho}+\bar{P})v}{2M_{{\rm Pl}}^{2}Hv}\,.\label{eqn:dbarN/dN_pre}
\end{align}
By substituting (\ref{eqn:dv/dphi}) to (\ref{eqn:dbarN/dN_pre}) and then solving it with respect to $d\bar{\mathcal{N}}/d\mathcal{N}$, one obtains 
\begin{equation}
\frac{d\bar{\mathcal{N}}}{d\mathcal{N}}=\frac{(\Gamma_{0}+3\Gamma_{1})\bar{\rho}}{\Gamma_{0}\bar{\rho}-3\Gamma_{1}\bar{P}}\left\{ 1-\frac{1}{2}\frac{1}{\Gamma_{0}+3\Gamma_{1}}\,\frac{d\ln\Theta}{d\mathcal{N}}+\frac{3\Gamma_{1}}{\Gamma_{0}+3\Gamma_{1}}\left[\frac{(\rho+P)+(\bar{\rho}+\bar{P})v}{\bar{\rho}v}\frac{\phi}{3H}+\frac{\rho+P}{\bar{\rho}v}\right]\right\}\,, \label{eqn:dbarN/dN}
\end{equation}
provided that $\Gamma_{0}\bar{\rho}-3\Gamma_{1}\bar{P}\neq0$.

We are now ready to reconstruct $v$ and the general (and as yet unspecified) observable $\Theta$, and thus $f_{0}$ and $f_{1}$, as functions of $\phi$. We assume that $\Theta$, $H$, $\rho$ and $P$ are given as functions of $\mathcal{N}$ and that $\bar{\rho}$ and $\bar{P}$ are given as functions of $\bar{\mathcal{N}}$. Then (\ref{eqn:dphidN}) and (\ref{eqn:dbarN/dN}), where $v$ is given by (\ref{eqn:v}), are differential equations for $\phi(\mathcal{N})$ and $\bar{\mathcal{N}}(\mathcal{N})$. Provided that $(\rho+P)+(\bar{\rho}+\bar{P})v>0$ and that $H>0$, (\ref{eqn:dphidN}) implies that $\phi(\mathcal{N})$ is an increasing function of $\mathcal{N}$ and thus admits the unique inverse function $\mathcal{N}(\phi)$. One then completes the reconstruction as 
\begin{equation}
v=\frac{M_{{\rm Pl}}^{2}\phi^{2}-3\rho(\mathcal{N}(\phi))}{3\bar{\rho}(\bar{\mathcal{N}}(\mathcal{N}(\phi)))}\,,\quad\Theta=\Theta(\mathcal{N}(\phi))\,.\label{eqn:v-reconstructed}
\end{equation}
Finally, $f_{0}$ and $f_{1}$ are given as functions of $\phi$, once (\ref{eqn:v-reconstructed}) is substituted to (\ref{eqn:f0f1-geffv}). This reconstruction method is quite general, as we do not specify which observable $\Theta$ we are looking at, just assuming it is a power law function of $f_{0}$ and $f_{1}$. 

In the remaining part of this paper we will focus on making this model suitable to describe weak gravity for dark matter at late times. Since weak gravity at short scales (i.e.\ for high values of $k$) will be affected mostly by a dust component, we will study since now on the case for which $\bar{\rho}=\bar{\rho}_{\rm cdm}+\Mpl^2\bar{\Lambda}$ and $\bar{P}=\bar{P}_{\rm cdm}-\Mpl^2\bar{\Lambda}$ with $\bar{P}_{\rm cdm}$=0, that is we model dark matter as a pressure-less fluid in the canonically transformed frame. In this case, as we shall see later on, the constants $\Gamma_{0}$ and $\Gamma_{1}$ will be both equal to unity for $\Theta = G_{\rm eff}/G_{N}$.

\section{Perturbation equations of motion}
\label{sec:perturbation}

In this section we derive perturbation equations of motion, adopting the VCCDM Lagrangian (\ref{eqn:VCCDM}). 

\subsection{Perturbation variables}

We introduce the following perturbation variables. For the metric we have
\begin{eqnarray}
ds_{3}^{2} & = & [a^{2}\,(1+2\zeta)\,\delta_{ij}+\partial_{i}\partial_{j}E]\,dx^{i}dx^{j}\,,\\
N & = & N(t)\,(1+\alpha)\,,\\
N_{i} & = & N(t)\,\partial_{i}\chi\,.
\end{eqnarray}
Later on we will set a 3D gauge as $E=0$. For the two 3D scalars we write them as
\begin{eqnarray}
\phi & = & \phi(t)+\delta\phi\,,\\
\lambda & = & \lambda(t)+\delta\lambda\,,
\end{eqnarray}
and for the gauge-fixing vector, we set
\begin{equation}
\lambda_{{\rm gf}}^{i}=\frac{1}{a^{2}}\,\delta^{ij}\partial_{j}\delta\lambda_{{\rm gf}}\,.
\end{equation}
The gauge fixing term equation of motion immediately gives
\begin{equation}
\delta\phi=0\,,
\end{equation}
which will be used in this whole section. In particular this leads
to $\delta(f_{0})=0$.

We consider $\bar{\rho}$ as consisting of two components, namely $\bar{\rho}_{{\rm cdm}}$ (for a dust component with its pressure $\bar{P}_{\rm cdm}=0$) and $\Mpl^{2}\,\tilde{\Lambda}$ (for a cosmological constant component with its pressure $-\Mpl^{2}\,\tilde{\Lambda}$). For the former, we define
\begin{equation}
\rho_{c}\equiv\bar{\rho}_{{\rm cdm}}\,v\,,
\end{equation}
for this combination at the level of the background ($\propto 1/(f_{1}a^{3})$) enters the modified Friedmann equation, Eq.\ (\ref{eqn:V}), and it represents, in the standard matter frame, the effective energy density of the dark component. We can now introduce the perturbation variables for the dark matter component as
\begin{equation}
\frac{\delta\rho_{c}}{\rho_{c}(t)}=\frac{1}{\rho_{c}(t)}\,\delta\!\left(\frac{\bar{\rho}_{{\rm cdm}}}{f_{1}f_{0}^{3/2}}\right)=\frac{\delta\bar{\rho}_{{\rm cdm}}}{\bar{\rho}_{{\rm cdm}}(t)}-\left[\frac{f_{1,\phi}}{f_{1}}+\frac{3}{2}\,\frac{f_{0,\phi}}{f_{0}}\right]\delta\phi\,,
\end{equation}
which, under a gauge transformation changes into
\begin{eqnarray}
\Delta\!\left(\frac{\delta\rho_{c}}{\rho_{c}(t)}\right) & = & \frac{1}{\rho_{c}(t)}\,\Delta(\delta\rho_{c})=-\frac{1}{\rho_{c}(t)}\,\frac{d}{dt}\!\left(\frac{\bar{\rho}_{{\rm cdm}}(t)}{f_{1}(t)f_{0}(t)^{3/2}}\right)\frac{\epsilon^{0}}{N}=-\frac{1}{\rho_{c}(t)}\,\frac{d}{dt}\!\left(\rho_{c}(t)\right)\frac{\epsilon^{0}}{N}\nonumber \\
 & = & -a^{3}\,f_{1}\,\frac{d}{dt}\!\left(\frac{1}{a^{3}\,f_{1}}\right)\frac{\epsilon^{0}}{N}=\left[3H+\frac{f_{1,\phi}}{f_{1}}\,\frac{\dot{\phi}}{N}\right]\epsilon^{0}\,.
\end{eqnarray}
On shell, one also has $\delta\phi=0$, so that
\begin{equation}
\frac{\delta\rho_{c}}{\rho_{c}(t)}=\frac{\delta\bar{\rho}_{{\rm cdm}}}{\bar{\rho}_{{\rm cdm}}(t)}\,.
\end{equation}
Finally for the dark matter component we also need to introduce
\begin{equation}
\sigma=\sigma(t)+\delta\sigma\,.
\end{equation}

In what follows we also consider the presence of standard model matter fields all being defined in the standard matter frame. Then the standard matter variables, both at the level of the background and perturbations, are introduced by means of the stress-energy tensor as defined in (\ref{eqn:Tmunu-sm}).

\subsection{Basic equations}

First of all let us consider the momentum constraint, that is the equation of motion $E_{\chi}=0$, found by taking a variation of the action for the perturbations with respect to the field $\chi$. We find it can be written as
\begin{equation}
E_{\chi}=\frac{\dot{\zeta}}{N}-H\alpha+\frac{1}{2}\,\delta\lambda-\frac{1}{2\Mpl^{2}}\,\sum_{I}n_{I}\rho_{I,n}v_{I}+\frac{1}{2\Mpl^{2}}\,f_{1}\rho_{c}\,\delta\sigma=0\,,
\end{equation}
so that it is possible and seems natural to define a new field $v_{c}$, the perturbation representing the speed of dark matter component, as to satisfy
\begin{eqnarray}
E_{\chi} & = & \frac{\dot{\zeta}}{N}-H\alpha+\frac{1}{2}\,\delta\lambda-\frac{1}{2\Mpl^{2}}\,(\rho_{I}+P_{I})\,v_{I}-\frac{1}{2\Mpl^{2}}\,\rho_{c}(t)\,v_{c}=0\,,
\end{eqnarray}
which leads to
\begin{equation}
\delta\sigma=-\frac{v_{c}}{f_{1}}\,.
\end{equation}
On performing a coordinate-gauge transformation, one finds
\begin{equation}
\Delta(v_{c})=-f_{1}\,\Delta(\delta\sigma)=f_{1}\,\frac{\dot{\sigma}}{N}\,\epsilon^{0}=\epsilon^{0}\,.
\end{equation}

Let us now study the matter equations of motion for the perturbations. On considering the equation of motion for the field $\delta\rho_{c}/\rho_{c}$, we find
\begin{equation}
-\frac{E_{\delta_{c}}}{\mu_{0}\mathsf{N}_{0}N}=\frac{\dot{v}_{c}}{N}+\alpha-\frac{f_{1,\phi}}{f_{1}}\,\frac{\dot{\phi}}{N}\,v_{c}=0\,,
\end{equation}
whereas for a cold baryonic fluid one gets the standard relation
\begin{equation}
\frac{E_{\delta_{b}}}{-Na^{3}\rho_{b}}=\frac{\dot{v}_{b}}{N}+\alpha=0\,,
\end{equation}
so that in general one finds $v_{c}\neq v_{b}$. This implies that, even on setting them equal as an initial condition, i.e.\ implementing adiabatic initial conditions, they will diverge at any later time, in general. In fact, one has that
\[
\frac{\dot{v}_{c}}{N}-\frac{\dot{v}_{b}}{N}=\frac{f_{1,\phi}}{f_{1}}\left(\frac{3}{2}\frac{1}{\Mpl^{2}}\,\sum_{I}(\rho_{I}+P_{I})+\frac{3}{2}\frac{1}{\Mpl^{2}}\,\rho_{c}\right)v_{c}\,. 
\]
We can rewrite the equation of motion for $v_{c}$ as
\begin{equation}
E_{v_{c}}=\frac{1}{N}\,\frac{d}{dt}\!\left(\frac{\delta\rho_{c}}{\rho_{c}}\right)+\frac{3}{N}\,\dot{\zeta}+\frac{k^{2}}{a^{2}}\,\chi-\frac{k^{2}}{a^{2}}\,\frac{f_{0}}{f_{1}^{2}}\,v_{c}=0\,,
\end{equation}
whereas for a baryonic fluid one finds
\begin{equation}
E_{v_{b}}=\frac{1}{N}\,\frac{d}{dt}\!\left(\frac{\delta\rho_{b}}{\rho_{b}}\right)+\frac{3}{N}\,\dot{\zeta}+\frac{k^{2}}{a^{2}}\,\chi-\frac{k^{2}}{a^{2}}\,v_{b}=0\,,
\end{equation}
so that we also find, in general, $\frac{\delta\rho_{b}}{\rho_{b}}\neq\frac{\delta\rho_{c}}{\rho_{c}}$, even when we set them equal at some initial time, because
\begin{equation}
\frac{1}{N}\,\frac{d}{dt}\!\left(\frac{\delta\rho_{c}}{\rho_{c}}\right)-\frac{1}{N}\,\frac{d}{dt}\!\left(\frac{\delta\rho_{b}}{\rho_{b}}\right)=\frac{k^{2}}{a^{2}}\left(\frac{f_{0}}{f_{1}^{2}}\,v_{c}-v_{b}\right).
\end{equation}

We also have that the equation of motion for the lapse perturbation gives
\begin{equation}
E_{\alpha}=3\,H\,\frac{\dot{\zeta}}{N}-3H^{2}\alpha+\frac{k^{2}}{a^{2}}\,H\,\chi+\frac{k^{2}}{a^{2}}\,\zeta+\frac{3}{2}\,H\,\delta\lambda-\frac{1}{2\Mpl^{2}}\sum_{I}\rho_{I}\left(\frac{\delta\rho_{I}}{\rho_{I}}\right)-\frac{1}{2\Mpl^{2}}\,\rho_{c}\left(\frac{\delta\rho_{c}}{\rho_{c}}\right)=0\,,
\end{equation}
giving once more a correct interpretation of the matter variables in terms of the perturbed Einstein equations. 

We can reduce the action by integrating out all the auxiliary fields leaving only the independent and propagating degrees of freedom, as to extract the no-ghost conditions, the speed of propagation, and the dynamics of the independent fields. Then we find for high $k$, the following no-ghost condition
\begin{equation}
\frac{f_{1}}{f_{0}}>0\,,\qquad\rho_{I}+P_{I}>0\,,
\end{equation}
whereas for the speed of propagation one finds
\begin{eqnarray}
c_{s,c}^{2} & = & 0\,,\\
c_{s,b}^{2} & = & 0\,,\\
c_{s,r}^{2} & = & \frac{1}{3}\,,
\end{eqnarray}
which coincide with the standard results, so that no ghost or small-scales instability is present.

\subsection{Newtonian independent variables dynamics}

In the following, we will consider the subcase of two fluids, both being pressureless, one being the CCDM, and the other representing a standard baryonic component. This is motivated by the fact that we want to investigate the implication of CCDM in the context of structure formation. For this aim we neglect any radiation component, and we find it convenient to introduce a Newtonian field redefintion. In particular, after having set the three-dimensional gauge for the scalar sector as $E=0$, we can introduce the following Newtonian time-gauge-invariant combinations
\begin{eqnarray}
\delta_{c} & \equiv & \frac{\delta\rho_{c}}{\rho_{c}}+\frac{\dot{\rho}_{c}}{N\rho_{c}}\,\chi\,,\\
\delta_{b} & \equiv & \frac{\delta\rho_{b}}{\rho_{b}}+\frac{\dot{\rho}_{b}}{N\rho_{b}}\,\chi\,,\\
\psi & \equiv & \alpha+\frac{\dot{\chi}}{N}\,,\\
\phi & \equiv & -\zeta-H\,\chi\,,
\end{eqnarray}
so that, in the high-$k$ limit the reduced Lagrangian density for the perturbations reduces to
\begin{equation}
\mathcal{L}=\frac{Na^{3}}{2}\,\frac{a^{2}}{k^{2}}\left[\frac{f_{1}^{2}}{f_{0}}\,\rho_{c}\,\frac{\dot{\delta}_{c}^{2}}{N^{2}}+\rho_{b}\,\frac{\dot{\delta}_{b}^{2}}{N^{2}}+\frac{\rho_{c}^{2}\delta_{c}^{2}}{2\Mpl^{2}}+\frac{\rho_{b}^{2}\delta_{b}^{2}}{2\Mpl^{2}}+\frac{\rho_{b}\rho_{c}\delta_{b}\delta_{c}}{\Mpl^{2}}\right].
\end{equation}
For this reduced Lagrangian, which implies that the system possesses two independent degrees of freedom, as expected, we can find the following equation of motion for the CCDM component
\begin{eqnarray*}
E_{c} & \equiv & -\partial_{t}\!\left[Na^{3}\,\frac{a^{2}}{k^{2}}\,\frac{f_{1}^{2}}{f_{0}}\,\rho_{c}\,\frac{\dot{\delta}_{c}}{N^{2}}\right]+Na^{3}\,\frac{a^{2}}{k^{2}}\frac{\rho_{c}^{2}}{2\Mpl^{2}}\,\delta_{c}+\frac{Na^{3}}{2}\,\frac{a^{2}}{k^{2}}\frac{\rho_{b}\rho_{c}}{\Mpl^{2}}\,\delta_{b}\\
 & = & -NH\partial_{\mathcal{N}}\!\left[\frac{a^{5}}{k^{2}}\,\frac{f_{1}^{2}}{f_{0}}\,\rho_{c}\,H\delta'_{c}\right]+Na^{3}\,\frac{a^{2}}{k^{2}}\frac{\rho_{c}^{2}}{2\Mpl^{2}}\,\delta_{c}+\frac{Na^{3}}{2}\,\frac{a^{2}}{k^{2}}\frac{\rho_{b}\rho_{c}}{\Mpl^{2}}\,\delta_{b}=0\,,
\end{eqnarray*}
where we have introduced the e-fold time variable, $\mathcal{N}=\ln(a/a_{0})$.
A similar equation holds for the baryonic component. Finally we can
write down the dynamical equation for both the pressureless components
as
\begin{eqnarray}
\delta''_{c} & + & \frac{\partial_{\mathcal{N}}\!\left[Ha^{2}\,\frac{f_{1}}{f_{0}}\right]}{Ha^{2}\,\frac{f_{1}}{f_{0}}}\,\delta'_{c}-\frac{f_{0}}{f_{1}^{2}}\,\frac{\rho_{c}}{2\Mpl^{2}H^{2}}\,\delta_{c}-\frac{f_{0}}{f_{1}^{2}}\,\frac{\rho_{b}}{2\Mpl^{2}H^{2}}\,\delta_{b}=0\,,\label{eq:d2dc}\\
\delta''_{b} & + & \frac{\partial_{\mathcal{N}}\!\left[Ha^{2}\right]}{Ha^{2}}\,\delta'_{b}-\frac{\rho_{c}}{2\Mpl^{2}H^{2}}\,\delta_{c}-\frac{\rho_{b}}{2\Mpl^{2}H^{2}}\,\delta_{b}=0\,.\label{eq:d2db}
\end{eqnarray}
On introducing $G_N=(8\pi\Mpl^2)^{-1}$, (\ref{eq:d2db}) is rewritten as
\[
\delta''_{b} + \frac{\partial_{\mathcal{N}}\!\left[Ha^{2}\right]}{Ha^{2}}\,\delta'_{b}
 - \frac{4\pi G_N}{H^2}(\rho_c\delta_c+\rho_b\delta_b)=0\,.
\]
Along the same line, the coefficients of the last two terms in (\ref{eq:d2dc}) define the effective gravitational constant $G_{\rm eff}$ for dark matter as a function of time,
\begin{equation}
4\pi G_{\rm eff}\equiv\frac{f_{0}}{f_{1}^{2}}\,\frac{1}{2\Mpl^{2}}\,,\qquad{\rm or}\qquad\frac{G_{\rm eff}}{G_{N}}=\frac{f_{0}}{f_{1}^{2}}\,,
\end{equation}
so that (\ref{eq:d2dc}) is rewritten as
\[
 \delta''_{c} + \frac{\partial_{\mathcal{N}}\!\left[Ha^{2}\,\frac{f_{1}}{f_{0}}\right]}{Ha^{2}\,\frac{f_{1}}{f_{0}}}\,\delta'_{c} - \frac{4\pi G_{\rm eff}}{H^2}(\rho_c\delta_c+\rho_b\delta_b)=0\,.
\]

Notice that the variables 
\begin{equation}
\delta_{m}\equiv\frac{\rho_{c}\delta_{c}+\rho_{b}\delta_{b}}{\rho_{c}+\rho_{b}}\,,\qquad{\rm and}\qquad v_{m}\equiv\frac{\rho_{c}v_{c}+\rho_{b}v_{c}}{\rho_{c}+\rho_{b}}\,,
\end{equation}
will both feel modifications due to the fact that $\frac{f_{0}}{f_{1}^{2}}\neq1$. It is also clear that, even imposing initial conditions for which $\delta_{c}(\mathcal{N}_{i})=\delta_{b}(\mathcal{N}_{i})$, and $v_{c}(\mathcal{N}_{i})=v_{b}(\mathcal{N}_{i})$, at any time after that in general one finds $\delta_{m}\neq\delta_{c}\neq\delta_{b}$, and $v_{m}\neq v_{c}\neq v_{b}$. In GR, instead, as long as we can trust the linear perturbation theory, we have that adiabatic initial conditions lead to $\delta_{m}=\delta_{c}=\delta_{b}$, and $v_{m}=v_{c}=v_{b}$. These differences in the dynamics should leave imprints on several observables related to the structure formation which will be discussed elsewhere.

Anyhow, demanding less structure formation at late times would lead in general to 
\begin{equation}
\frac{f_{0}}{f_{1}^{2}}<1\,,\qquad\textrm{at late times}.
\end{equation}
However, it should be noted that also the friction term gets modifications, which will in turn modify the dynamics of the evolution of structure formation, therefore a numerical detailed study is needed to understand the observables for these theory and its constraints coming from late-time data.

Finally then we have
\begin{eqnarray}
V(\phi) & = & \frac{\tilde{\Lambda}}{f_{1}f_{0}^{3/2}}\,,\\
\frac{G_{\rm eff}}{G_{N}} & = & \frac{f_{0}}{f_{1}^{2}}\,,
\end{eqnarray}
so that out of these two functions one can achieve two kinds of modifications, one for the background and the other for matter perturbations.

On the other hand we also find in the high-$k$ approximation, that
\begin{eqnarray}
\psi & = & -\phi=-\frac{a^{2}}{k^{2}}\,\frac{1}{2\Mpl^{2}}\,(\rho_{b}\delta_{b}+\rho_{c}\delta_{c})+\mathcal{O}(k^{-4})\,,\\
\chi & = & \mathcal{O}(k^{-4})\,,\\
v_{b} & = & \frac{a^{2}}{k^{2}}\,\frac{\dot{\delta}_{b}}{N}+\mathcal{O}(k^{-4})\,,\\
v_{c} & = & \frac{a^{2}}{k^{2}}\,\frac{f_{1}^{2}}{f_{0}}\,\frac{\dot{\delta}_{c}}{N}+\mathcal{O}(k^{-4})\,,
\end{eqnarray}
which lead to interesting consequences. In particular, this theory of gravity does not introduce any effective shear component, as $\psi+\phi=0$, as in GR. Furthermore the Poisson equation does not change the form, but the couplings in the dark matter equation do. Therefore the evolution  of $\delta$'s will in turn modify the dynamics of $\psi$ (or $\phi$) with respect to the one of GR.

\subsection{Simple example}

\begin{figure}
\includegraphics[width=13cm]{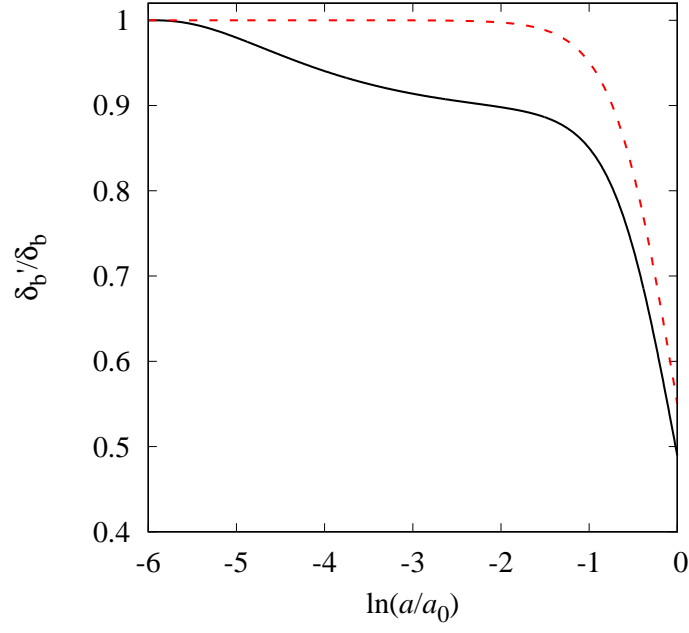}
\caption{In this figure we plot the behavior of $\delta'_{b}/\delta_{b}$ for the $\Lambda$CDM (dotted red line) and for the VCCDM (continuous black line). We use the same GR initial conditions at early times for both theories, namely $\delta_{b}(\mathcal{N}=-6)=\delta_{c}(\mathcal{N}=-6)=a_{i}$, and $\delta_{b}'(\mathcal{N}=-6)=\delta_{c}'(\mathcal{N}=-6)=a_{i}$. We have also imposed conditions for $\Omega_{c}=\rho_{c}/(3\Mpl^{2}H^{2})$ and $\Omega_{b}=\rho_{b}/(3\Mpl^{2}H^{2})$ so that today we have $\Omega_{c0}=0.3$ and $\Omega_{b0}=0.04$. For the VCCDM model we have fixed $G_{\rm eff}/G_{N}=0.8$.\label{fig:fs8}}
\end{figure}

Just to give an example of what the phenomenology of this theory could lead to, we solve Eqs.\ (\ref{eq:d2dc}) and (\ref{eq:d2db}), for a simple case, namely when both $f_{0}$ and $f_{1}$ are constants. We consider $\bar{\rho}=\bar{\rho}_{c}+\Mpl^2\bar{\Lambda}$ and $\rho=\rho_{b}$. In this case we have that
\begin{eqnarray}
 \rho_{c} & = & \frac{\bar{\rho}_{c}}{f_{1}f_{0}^{3/2}}=\frac{\bar{\rho}_{c0}}{f_{1}f_{0}^{3/2}}\left(\frac{\bar{a}_{0}}{\bar{a}}\right)^{3}=\rho_{c0}\left(\frac{a_{0}}{a}\right)^{3}\,,\\
 \Mpl^2 V & = & \frac{\Mpl^{2}\,\tilde{\Lambda}}{f_{1}f_{0}^{3/2}}={\rm const.}
\end{eqnarray}
In this case we also find that $\phi=-3H$, and we obtain the standard $\Lambda$CDM background evolution. However, matter perturbations do depend on the extra parameter $G_{\rm eff}/G_{N}$, which is a constant and differs from unity in general. The results are shown in Fig.\ \ref{fig:fs8}. The result shows that a weaker effective gravitational constant for $\delta_{c}$ leads to a smaller growth rate also for $\delta_{b}$.

In this same case, we notice that
\begin{equation}
\frac{\dot{v}_{c}}{N}-\frac{\dot{v}_{b}}{N}=0\,,\qquad{\rm or}\qquad v_{c}=v_{b}+\frac{f_{1}^{2}}{f_{0}}\,\frac{A(k)}{k^{2}}\,,
\end{equation}
which leads to
\begin{equation}
\frac{a^{2}}{k^{2}}\,\frac{f_{1}^{2}}{f_{0}}\,\frac{\dot{\delta}_{c}}{N}=\frac{a^{2}}{k^{2}}\,\frac{\dot{\delta}_{b}}{N}+\frac{f_{1}^{2}}{f_{0}}\,\frac{A(k)}{k^{2}}\,,\label{eq:dotDcDb}
\end{equation}
and
\begin{eqnarray}
\delta_{c} & = & \frac{f_{0}}{f_{1}^{2}}\,\delta_{b}+A(k)\int^{t}N(t')/a(t')^{2}dt'+B(k)\,.\\
 & = & \frac{f_{0}}{f_{1}^{2}}\,\delta_{b}+A(k)\int_{\mathcal{N}_{i}}^{\mathcal{N}}\frac{1}{Ha^{2}}\,d\mathcal{N}'+B(k)\,.
\end{eqnarray}
We could get the same result found above by combining Eqs.\ (\ref{eq:d2dc}) and (\ref{eq:d2db}) to get 
\begin{equation}
\delta''_{c}+\frac{\partial_{\mathcal{N}}\!\left[Ha^{2}\right]}{Ha^{2}}\,\delta'_{c}=\frac{f_{0}}{f_{1}^{2}}\left[\delta''_{b}+\frac{\partial_{\mathcal{N}}\!\left[Ha^{2}\right]}{Ha^{2}}\,\delta'_{b}\right],
\end{equation}
which is indeed solved by 
\begin{equation}
\delta'_{c}=\frac{f_{0}}{f_{1}^{2}}\,\delta'_{b}+\frac{A(k)}{Ha^{2}}\,,
\end{equation}
which is equivalent to Eq.\ (\ref{eq:dotDcDb}). The background equation of motion reads
\begin{equation}
\frac{\dot{H}}{N}=-4\pi G_{N}(\rho_{b}+\rho_{c})\,,\qquad{\rm or}\qquad\frac{H'}{H}=-\frac{3}{2}\,(\Omega_{b}+\Omega_{c})\,,
\end{equation}
so that we have
\begin{eqnarray}
\delta''_{b} & + & \left[2-\frac{3}{2}\,(\Omega_{b}+\Omega_{c})\right]\delta'_{b}-\frac{\rho_{b}}{2\Mpl^{2}H^{2}}\left(1+\frac{f_{0}}{f_{1}^{2}}\,\frac{\rho_{c}}{\rho_{b}}\right)\delta_{b}=\frac{\rho_{c}}{2\Mpl^{2}H^{2}}\left[A(k)\int_{\mathcal{N}_{i}}^{\mathcal{N}}\frac{1}{Ha^{2}}\,d\mathcal{N}'+B(k)\right]\,,
\end{eqnarray}
and, for the initial conditions
\begin{equation}
\delta_{c}(\mathcal{N}_{i})=\delta_{b}(\mathcal{N}_{i})=\delta'_{c}(\mathcal{N}_{i})=\delta'_{b}(\mathcal{N}_{i})=a_{i},
\end{equation}
we find
\begin{eqnarray}
\delta''_{b} & + & \left[2-\frac{3}{2}\,(\Omega_{b}+\Omega_{c})\right]\delta'_{b}-\frac{3}{2}\left(\Omega_{b}+\frac{f_{0}}{f_{1}^{2}}\,\Omega_{c}\right)\delta_{b}=\frac{3a_{i}\Omega_{c}}{2}\left(1-\frac{f_{0}}{f_{1}^{2}}\right)\left[\int_{\mathcal{N}_{i}}^{\mathcal{N}}\frac{H_{i}a_{i}^{2}}{Ha^{2}}\,d\mathcal{N}'+1\right].
\end{eqnarray}
In order to understand the results qualitatively, since $0<\rho_{b}/\rho_{c}=\Omega_{b0}/\Omega_{c0}\ll1$ and assuming $\delta_{b}\sim\delta_{c}$, we have
\begin{eqnarray}
\delta''_{c} & = & -\frac{\partial_{\mathcal{N}}\!\left[Ha^{2}\right]}{Ha^{2}}\,\delta'_{c}+\frac{f_{0}}{f_{1}^{2}}\,\frac{\rho_{c}\delta_{c}}{2\Mpl^{2}H^{2}}\left[1+\frac{\rho_{b}\delta_{b}}{\rho_{c}\delta_{c}}\right]\nonumber \\
 & \approx & -\frac{\partial_{\mathcal{N}}\!\left[Ha^{2}\right]}{Ha^{2}}\,\delta'_{c}+\frac{3}{2}\,\frac{f_{0}}{f_{1}^{2}}\,\Omega_{c}\,\delta_{c}\,,
\end{eqnarray}
whose solution will be suppressed compared to GR. In turn, this reduced $\delta_{c}$ induces a lower value for $\delta_{b}$, since
\begin{eqnarray}
\delta''_{b} & = & -\frac{\partial_{\mathcal{N}}\!\left[Ha^{2}\right]}{Ha^{2}}\,\delta'_{b}+\frac{\rho_{c}\delta_{c}}{2\Mpl^{2}H^{2}}\left[1+\frac{\rho_{b}\delta_{b}}{\rho_{c}\delta_{c}}\right]\nonumber \\
 & \approx & -\frac{\partial_{\mathcal{N}}\!\left[Ha^{2}\right]}{Ha^{2}}\,\delta'_{b}+\frac{3}{2}\,\Omega_{c}\,\delta_{c}\,.
\end{eqnarray}

\section{Conclusions}\label{sec:concl}

We have proposed a new cosmological framework that can weaken gravity for dark matter at late time without introducing extra gravitational degrees of freedom. We have achieved this goal by integrating dark matter into a recently proposed type-II minimally modified gravity (MMG) theory that can mimic any dark energy component at the level of the homogeneous and isotropic cosmological background. Concretely, we have introduced a dark matter Lagrangian and a cosmological constant in a canonically transformed frame of general relativity (GR). In order to avoid extra degrees of freedom in the physical phase space, before adding both the dark matter and the cosmological constant, we have actually imposed a gauge-fixing constraint that splits the first-class constraint associated with the temporal diffeomorphism into a pair of second-class constraints. After going back to the original frame by the inverse canonical transformation, we have Legendre transformed the Hamiltonian to obtain the action of the theory in the original frame, where one can add the standard matter fields. The total action in this framework is given by (\ref{eqn:totalaction-general}), where the dark matter is coupled to the effective metric (\ref{eqn:4dmetric-dark}) and $V(\phi)$ plays the role of dark energy. 

In the standard matter frame, the dynamics of both the dark matter and cosmological constant components does not coincide with the usual GR one. In particular, we can use the two free functions in order to reconstruct the dynamics, virtually any, of two chosen background dependent observables (i.e.\ scale independent), e.g.\ $H(z)$ and $G_{\rm eff}(z)/G_N$. We have developed a general and concrete reconstruction method in section \ref{sec:reconstruction}. With this new cosmological framework, we thus believe that it should be possible to fit today's experimental data sets, which turn out to be puzzling in the standard framework. This expectation comes from the fact that we do not introduce any new physical degree of freedom (besides the one corresponding to the dark matter) and, as such, the model does not need to obey any extra ghost or gradient instability conditions in addition to GR. 

The large freedom of setting the dynamics of two background observables that we have in this theory is therefore due to the absence of extra gravitational degrees of freedom and, in turn, this is a consequence of the fact that the four-dimensional diffeomorphism invariance is broken to the three-dimensional spatial diffeomorphism invariance in the gravity sector. 

We conclude that the model presented here explicitly shows that there exist theories which can have a higher $H_0$ and, at the same time, weaker gravity, compared to $\Lambda$CDM, still keeping cosmology stable and ghost free.

\begin{acknowledgments}
The work of A.D.F.\ was supported by Japan Society for the Promotion of Science Grants-in-Aid for Scientific Research No.\ 20K03969. The work of S.M.\ was supported in part by Japan Society for the Promotion of Science Grants-in-Aid for Scientific Research No.~17H02890, No.~17H06359, and by World Premier International Research Center Initiative, MEXT, Japan. 
\end{acknowledgments}

\appendix

\section{Dust action in GR}

The scalar dust action can be written as follows
\begin{equation}
S=-\frac{1}{2}\int d^{4}x\sqrt{-g}\,\rho\,(\partial_{\mu}\sigma\partial_{\nu}\sigma\,g^{\mu\nu}+1)\,.
\end{equation}
After the $3+1$ decomposition, we find the following Lagrangian density
\begin{equation}
\mathcal{L}=-\frac{1}{2}\,N\,\sqrt{\gamma}\,\rho\left[-\frac{\dot{\sigma}^{2}}{N^{2}}+2\,\frac{N^{i}}{N^{2}}\,\dot{\sigma}\nabla_{i}\sigma+\left(\gamma^{ij}-\frac{N^{i}N^{j}}{N^{2}}\right)\nabla_{i}\sigma\nabla_{j}\sigma+1\right]\,.
\end{equation}
The relation between $\pi_{\sigma}$ and $\dot{\sigma}$ is
\begin{equation}
\dot{\sigma}=\frac{N}{\rho}\,\frac{\pi_{\sigma}}{\sqrt{\gamma}}+N^{i}\partial_{i}\sigma\,,
\end{equation}
so that the matter Hamiltonian alone becomes
\[
H_{{\rm m0}}=\int d^{3}x\left\{ \frac{N\sqrt{\gamma}}{2\rho}\,\left[\left(\frac{\pi_{\sigma}}{\sqrt{\gamma}}\right)^2+\rho^{2}\,(1+\partial_{i}\sigma\partial_{j}\sigma\,\gamma^{ij})\right]+\lambda_{\rho}\pi_{\rho}+\pi_{\sigma}N^{i}\partial_{i}\sigma\right\} .
\]
We have introduced here two fields, $\sigma$ and $\rho$. In general we should then study $H_{{\rm tot}0}$, including the gravity degrees of freedom, where
\begin{equation}
H_{{\rm tot}0}=N\mathcal{H}_{0}(\gamma,\pi)+N^{i}\mathcal{H}_{i}(\gamma,\pi)+H_{{\rm m}0}\,.
\end{equation}
Notice that this Hamiltonian gives a contribution to the total $N$-constraint and to the $N^{i}$-constraints. And, finally, it adds a new constraint $\pi_{\rho}\approx0$.

However, for a test function $v_{\rho}$, we have that the time-derivative of the $\pi_{\rho}$ constraint is determined only by the matter Hamiltonian as follows
\begin{equation}
\{v_{\rho}\pi_{\rho},H_{{\rm tot}0}\}=\{v_{\rho}\pi_{\rho},H_{{\rm m}0}\}\approx0\,,
\end{equation}
which leads to a secondary constraint
\begin{equation}
C_{\rho}=\sqrt{\gamma}\left[1-\left(\frac{\pi_{\sigma}}{\sqrt{\gamma}\rho}\right)^2+\partial_{i}\sigma\partial_{j}\sigma\,\gamma^{ij}\right]\approx0,
\end{equation}
so that
\begin{equation}
H_{{\rm m}}=\int d^{3}x\left\{ \frac{N\sqrt{\gamma}}{2\rho}\,\left[\left(\frac{\pi_{\sigma}}{\sqrt{\gamma}}\right)^2+\rho^{2}\,(1+\partial_{i}\sigma\partial_{j}\sigma\,\gamma^{ij})\right]+\lambda_{\rho}\pi_{\rho}+\pi_{\sigma}N^{i}\partial_{i}\sigma+\lambda_{2}\,C_{\rho}\right\}\,.
\end{equation}
Then it is also clear that for the updated $H_{{\rm tot}}=N\mathcal{H}_{0}(\gamma,\pi)+N^{i}\mathcal{H}_{i}(\gamma,\pi)+H_{{\rm m}}$, we have
\begin{eqnarray}
\{v_{\rho}\pi_{\rho},H_{{\rm tot}}\} & \approx & 0\,,\qquad\Rightarrow\qquad\lambda_{2}\approx0\,,\\
\{\sqrt{\gamma}\,v_{N}C_{N}^{{\rm lapse}},H_{{\rm tot}}\} & \approx & 0\,,\qquad{\rm automatically}\,,\\
\{\sqrt{\gamma}\,v^{i}C_{i}^{{\rm shift}},H_{{\rm tot}}\} & \approx & 0\,,\qquad{\rm automatically}\,,\\
\{\sqrt{\gamma}\,v_{2}C_{\rho},H_{{\rm tot}}\} & \approx & 0\,,\qquad\textrm{can be solved for \ensuremath{\lambda_{\rho}}}\,,
\end{eqnarray}
on the constraint surface. In GR, one can show that there are four 1st-class constraints (a linear combination of $C_{N}$ and $\bar{\pi}_{\rho}$, and generalized $C_{i}^{{\rm shift}}$), and two 2nd-class constraints, $\bar{\pi}_{\rho}$ and $C_{\rho}$. Hence we find $\frac{1}{2}(6\times2+1\times2+1\times2-4\times2-2\times1)=3$ degrees of freedom as expected.

\end{document}